\documentstyle[12pt]{article}

\def\hybrid{\topmargin -20pt    \oddsidemargin 0pt
        \headheight 0pt \headsep 0pt
        \textwidth 6.25in       
        \textheight 9.5in       
        \marginparwidth .875in
        \parskip 5pt plus 1pt   \jot = 1.5ex}

\hybrid

\catcode`\@=11

\def\marginnote#1{}
\newcount\hour
\newcount\minute
\newtoks\amorpm
\hour=\time\divide\hour by60
\minute=\time{\multiply\hour by60 \global\advance\minute by-\hour}
\edef\standardtime{{\ifnum\hour<12 \global\amorpm={am}%
        \else\global\amorpm={pm}\advance\hour by-12 \fi
        \ifnum\hour=0 \hour=12 \fi
        \number\hour:\ifnum\minute<10 0\fi\number\minute\the\amorpm}}
\edef\militarytime{\number\hour:\ifnum\minute<10 0\fi\number\minute}

\def\draftlabel#1{{\@bsphack\if@filesw {\let\thepage\relax
   \xdef\@gtempa{\write\@auxout{\string
      \newlabel{#1}{{\@currentlabel}{\thepage}}}}}\@gtempa
   \if@nobreak \ifvmode\nobreak\fi\fi\fi\@esphack}
        \gdef\@eqnlabel{#1}}
\def\@eqnlabel{}
\def\@vacuum{}
\def\draftmarginnote#1{\marginpar{\raggedright\scriptsize\tt#1}}

\def\draft{\oddsidemargin -.5truein
        \def\@oddfoot{\sl 2nd draft \hfil
        \rm\thepage\hfil\sl\today\quad\militarytime}
        \let\@evenfoot\@oddfoot \overfullrule 3pt
        \let\label=\draftlabel
        \let\marginnote=\draftmarginnote
   \def\@eqnnum{(\theequation)\rlap{\kern\marginparsep\tt\@eqnlabel}%
\global\let\@eqnlabel\@vacuum}  }

\def\preprint{\twocolumn\sloppy\flushbottom\parindent 2em
        \leftmargini 2em\leftmarginv .5em\leftmarginvi .5em
        \oddsidemargin -.5in    \evensidemargin -.5in
        \columnsep .4in \footheight 0pt
        \textwidth 10.in        \topmargin  -.4in
        \headheight 12pt \topskip .4in
        \textheight 6.9in \footskip 0pt
        \def\@oddhead{\thepage\hfil\addtocounter{page}{1}\thepage}
        \let\@evenhead\@oddhead \def\@oddfoot{} \def\@evenfoot{} }

\def\numberbysection{\@addtoreset{equation}{section}
        \def\theequation{\thesection.\arabic{equation}}}

\def\underline#1{\relax\ifmmode\@@underline#1\else
        $\@@underline{\hbox{#1}}$\relax\fi}

\newskip\humongous \humongous=0pt plus 1000pt minus 1000pt
\def\caja{\mathsurround=0pt}
\def\eqalign#1{\,\vcenter{\openup1\jot \caja
        \ialign{\strut \hfil$\displaystyle{##}$&$
        \displaystyle{{}##}$\hfil\crcr#1\crcr}}\,}
\newif\ifdtup

\relax

\renewcommand{\theequation}{\arabic{equation}}

\newcommand{\rf}[1]{(\ref{#1})}
\newcommand{\be}{\begin{equation}}
\newcommand{\ee}{\end{equation}}
\newcommand{\bea}{\begin{eqnarray}}
\newcommand{\eea}{\end{eqnarray}}
\newcommand{\pa}{\partial} 
\begin{document}
 
\title{On Quantum T-duality in $\sigma$ models}

\author{
         J.~Balog$^{(1)}$, P.~Forg\'acs$^{(1)}$\thanks{
        corresponding author E-mail: forgacs@celfi.phys.univ-tours.fr}
        ,\\ 
N.~Mohammedi$^{(1)}$,  
 L.~Palla$^{(2)}$
 and J.~Schnittger$^{(1)}$ \\
       {\small $^{(1)}$ Laboratoire de Math. et Physique Theorique,} \\
       {\small CNRS UPRES-A 6083} \\
       {\small Universit\'e de Tours}\\
       {\small Parc de Grandmont, F-37200 Tours, France}\\
       {\small $^{(2)}$ Institute for Theoretical Physics,
        Roland E\"otv\"os University,} \\
       {\small  H-1088, Budapest, Puskin u. 5-7, Hungary }\\
}

\maketitle
\begin{abstract}

The problem of quantum equivalence between non-linear sigma models 
related
by Abelian or non-Abelian T-duality is studied in perturbation theory.
Using the anomalous Ward identity for Weyl symmetry we derive 
a relation between the Weyl anomaly coefficients of the original 
and dual theories. The analysis
is not restricted to conformally invariant backgrounds. 
The formalism is applied to the study of two examples. The first
is a model based on $SU(2)$ non-Abelian T duality. The second
represents a simple realization of Poisson-Lie T duality involving
the Drinfeld double based on $SU(2)$. In both cases 
quantum T duality is established at the 1-loop level.

\end{abstract}
\vspace{11 mm}
\begin{center}
key words: sigma models, duality, quantum corrections
\end{center}
\vfill\eject

\section{Introduction}

Target space duality (T-duality) symmetries play an important role
in modern string theory. A simple and elegant way to introduce 
T-duality is to identify them as canonical transformations \cite{can} 
in the $\sigma$-model formulation of string theory. The advantage of
the 
canonical transformation approach to T-duality is that it treats 
various kinds of T-duality transformations 
(Abelian \cite{Tab}, non-Abelian (NAD) \cite{OQ,AAL} and 
Poisson-Lie (PL) \cite{TPL,TU}) on an equal
footing.
Since the canonical transformation approach is 
classical, it leaves open the question whether 
 the duality related  $\sigma$-models give equivalent quantum field
 theories.
 
In quantum theory the usual way to argue for the equivalence between
the
dually related theories is based on gauging the isometries
in the functional integral representation. This applies to
Abelian 
and non-Abelian dualities, while it leaves open the problem of Poisson
- Lie 
duality, which does not require any isometry. For Conformal Field
Theories 
(CFT-s) it has been convincingly argued in 
\cite{RocVer} that the original
and dual
models are two different functional integral representations of the
same 
CFT. For some special cases (
ordinary and gauged Wess-Zumino-Witten-Novikov (WZWN) models) this
equivalence
has been shown explicitly on the level of current algebras.
Nevertheless an 
example was put forward in \cite{GRV} where two classically
equivalent, 
non-Abelian duality related conformal invariant sigma models were found
to be 
inequivalent at the one loop level. Subsequently the explanation of
this 
non - equivalence has been given \cite{AAL} by showing that the
gauging 
procedure is afflicted with anomalies, if the structure constants of
the 
gauge/isometry group have non vanishing traces.    

Conformal invariance is not a prerequisite for
the existence of quantum T-duality. Thus it is appropriate
to address this problem outside the context of string theory  
in the  general setting of renormalizable
sigma models, dropping the requirement of conformal symmetry. 
Therefore in our study we employ perturbative
techniques of quantum field theory, 
i.e. the loop expansion.

Quantum equivalence 
of dually related (renormalizable) sigma models 
is a non trivial 
problem already at the conceptual level. 
Any satisfactory criterion
should be based on
the comparison of physical quantities as opposed to just considering
beta functions.
If
there are
global symmetries in the model then their associated conserved
quantities 
(Noether currents) may be considered physical. 
The definition of
physical 
quantities, however, is not very clear in diffeomorphic invariant
sigma models without a sufficient 
number of isometries. Bearing this in mind we work in a
definite 
renormalization scheme (background field method with dimensional 
regularization), hence our results are - at least superficially -
dependent 
on this particular choice. On the other hand the quantities we compare
--the Weyl anomaly coefficients-- are at least free of field 
 redefinition and gauge transformation ambiguities \cite{ref15}.
They appear naturally and are physical in the sense that they describe
the trace of the 
world sheet energy momentum tensor.

The study of
examples so far has revealed no violations of quantum equivalence 
at the 1-loop level for Abelian or non-Abelian T-duality 
(apart from the class of anomalous models of the type of Ref.\ \cite{GRV}).
A general proof of the one loop equivalence  
for the {\sl Abelian case} has been given in \cite{HO}. At the
two-loop 
level, the situation is somewhat different. As shown in \cite{BFHP},
even
for the simplest examples \cite{Ts} T-duality can be maintained as an
 exact quantum symmetry only at the price of modifying the classical 
formulas \cite{Tab} accompanied by an appropriate change in the 
renormalization scheme.  
For the general Abelian case, 
a form of the 2-loop modifications has been determined in \cite{KM} 
through a consideration of the string effective action;
(though it remains to 
see whether they fit into the conventional field theoretical
renormalization scheme). 

Based on the study  of a number of examples \cite{ex}, we expect a 
similar pattern for non-Abelian T-duality. On the other hand,
there are too few examples worked
out \cite{plex} \cite{plex1} 
for the most general case of Poisson-Lie T-duality to
draw any conclusion at this stage.

The goal of the present paper is twofold: First, we want to establish
a general relation between the Weyl-anomaly 
coefficients (and beta functions) of the original model and those of
its dual in a general Abelian or non-Abelian
context. The argument is independent of the loop order
when applied to the
bare quantities, but we consider its implications in terms
of renormalized couplings only to 1 loop order. We establish it on
general grounds starting from the
assumption that the standard gauging procedure is valid
in the path integral\footnote{This 
excludes, in particular, the anomalous
cases of \cite{AAL}, \cite{GRV}.}.
 Our perspective is complementary to that of Ref.\ \cite{HO}
 where, - in the Abelian case, - such a relation was postulated, 
and verified by explicit calculation. 
This is the subject of section 2, where we point out as well that
the above relation also implies the 1-loop equivalence of the
renormalized partition functions. 

In the second half of the paper, we turn to the study of concrete
examples. In section 3, we illustrate our general considerations with
a non-Abelian example (based on $SU(2)$) with a single spectator field.
The beta functions/Weyl anomaly coefficients are explicitly computed,
 and we verify the validity of the relation mentioned above. 
In section 4, we analyse a simple example of Poisson-Lie type, 
involving the Drinfeld double based on $SU(2)$. 
Two new difficulties appear in the general PL setting: First, no
gauging procedure is available to date
\footnote{Note that the path integral derivation given by Tyurin and
von Unge, 
\cite{TU}, is not a conventional gauging.}, thus our starting
assumption of
Sect.\ 2 is lacking even a formal derivation in this case. 
Second, while one can easily show
that the isometries responsible for Abelian/non-Abelian duality
survive renormalization, it is less obvious that the criterion for 
PL dualizability does:  it is not a priori clear that the Poisson Lie 
dualizability of the bare fields is inherited by the renormalized
ones. 
Our general argument of section 2 is thus 
inapplicable.  We show, however,
 by explicit comparison of the one loop beta functions/Weyl anomaly 
coefficients of the dual partners that the 
relation derived in Section 2   
does hold for our example. This provides further evidence for the
existence of quantum PL T-duality.

\section{General considerations}

We will consider the most general 2-dimensional $\sigma$-model
with background metric $g_{\mu\nu}(X)$, torsion potential
$b_{\mu\nu}(X)$ and dilaton field $\phi(X)$. To simplify the
notation, sometimes we will use the generic symbol $w$ whose
components are the various background fields:
\bea
w^{(g)}_{\mu\nu}=g_{\mu\nu}\,,   \label{wcomponent1}\\
w^{(b)}_{\mu\nu}=b_{\mu\nu}\,,   \label{wcomponent2}\\
w^{(\phi)}=\phi\,.               \label{wcomponent3}
\eea
In renormalization theory the above finite fields will play the
r\^ole of the renormalized couplings. The analogous bare (unrenormalized)
fields will be denoted by the corresponding capital symbols
$G_{\mu\nu}$, $B_{\mu\nu}$ and $\Phi$ and collectively denoted
by $W$. The Lagrangian is written in terms of the bare fields as
\be
{\cal L}=\frac{1}{2}\gamma^{ab}
G_{\mu\nu}\partial_aX^\mu\partial_bX^\nu
+\frac{i}{2}\epsilon^{ab}
B_{\mu\nu}\partial_aX^\mu\partial_bX^\nu
+\alpha^\prime R^{(2)}\Phi\,.
\ee
Here $R^{(2)}$ is the worldsheet curvature and
the worldsheet metric $\gamma_{ab}$ can be brought to the
conformally flat form
\be
\gamma_{ab}(z)=e^{\sigma(z)}\delta_{ab}\,.
\ee
The partition function of the $\sigma$-model is obtained by
functionally integrating over the coordinate fields $X^\mu(z)$.
\be
Z[G,B,\Phi;\sigma]=\int[{\cal D}X^\mu]\,
e^{-S[G,B,\Phi;\sigma]}\,,
\label{Zbare}
\ee
where
\be
S[G,B,\Phi;\sigma]=\frac{1}{2\pi\alpha^\prime}
\int d^2z\sqrt{\gamma}\, {\cal L}\,.
\label{action}
\ee

The action \rf{action} is obviously invariant under
diffeomorphisms, i.e. transformations of the target
space coordinates $X^\mu$:
\be
S[W;\sigma]=S[W^{\cal D};\sigma]\,,
\label{diffeo}
\ee
where the diffeomorphism ${\cal D}$ can either be a finite
transformation, or, in perturbation theory (using dimensional regularization), 
a power series
in $\alpha^\prime$ with $\epsilon$ poles.
 
Similarly \rf{action} is invariant under the gauge transformation
\be
S[W;\sigma]=S[W^\Delta;\sigma]\,,
\label{gauge}
\ee
where
\be
B^\Delta_{\mu\nu}=B_{\mu\nu}+\partial_\mu\Delta_\nu-\partial_\nu\Delta_\mu
\label{Delta}
\ee
with $\Delta_\mu$ either finite or a perturbative series.
 
\subsection{Compatibility between duality transformations\\ 
and the Weyl flow}

The way we are going to check quantum T duality in the examples that
we will analyze is not by comparison of the partition functions
themselves, but rather of the Weyl anomaly coefficients, objects that
are closely related to  beta functions. 
 They are defined by the anomalous Ward identity
\be
\frac{\delta Y}{\delta\sigma(z)}=
\langle T^a_a(z)\rangle=
\langle{\cal L}(\overline\beta)\rangle\,,
\label{Weyl1}
\ee
Here, $Y[w;\sigma]$ denotes the renormalized partition function, 
related to the bare one of \rf{Zbare} by
\be
Y[w;\sigma]=Z[W(w);\sigma]\,.
\label{partrenorm}
\ee
Furthermore, $T_{ab}$ is the worldsheet energy-momentum tensor.
Its anomalous trace is of the form of the original
Lagrangian ${\cal L}$ where the Weyl anomaly coefficients
$\overline\beta(w)$ play the r\^ole of the background couplings
 \cite{ref15}. The Weyl anomaly coefficients differ from 
the $\beta$-functions
of the model only by additive shifts containing the dilaton (see below),
 but, unlike the $\beta$-functions, are free
of the ambiguities coming from diffeomorphisms and gauge
transformations.

In ref. \cite{HO}, the following relation connecting the Weyl anomaly
coefficients of the original model and its dual was proposed: 
\be
 \overline\beta( w)
\frac{\partial\Gamma}{\partial w}(w)=\overline\beta(\tilde w)
\,, \qquad \tilde w \equiv\Gamma(w)\,,
\label{Haag}
\ee
where $ \Gamma(w)$ denotes the duality transformations of the couplings. 
If $\hat T$ denotes the operator implementing the operation 
$w\to \tilde w$ and $\hat R$ the flow operator corresponding to
$$
w \ \ \to w+\lambda \bar\beta(w), 
$$
(where $\lambda$ is the infinitesimal flow parameter),  
then \rf{Haag} can be interpreted as the commutativity of the
``Weyl flow'' $\hat R$ with T duality, $[\hat T,\hat R]=0$. 
The authors of ref. \cite{HO}, who postulated \rf{Haag} somewhat
heuristically,
 proved that it is satisfied at least to 1 loop order 
for the case of Abelian duality.
Furthermore, they noted  that it holds exactly and not only up to 
diffeomorphisms and gauge transformations, as one may have expected, 
but did not provide any explanation for this fact. Our goal in this 
subsection will be to derive \rf{Haag} from a general point of view 
encompassing Abelian and non-Abelian T duality alike, and to explain the
absence of diffeomorphism and gauge terms. Within this setting, 
the function $\Gamma$ takes the general form
\bea
\tilde g_{\mu\nu}=\Gamma^{(g)}_{\mu\nu}(g,b)\,,  \\
\tilde b_{\mu\nu}=\Gamma^{(b)}_{\mu\nu}(g,b)\,,  \\
\tilde\phi=\phi+S(g,b)\,.
\label{dilshift}
\eea
For  Abelian and non-Abelian duality,  the set of T-dualizable
backgrounds is a linear space, since the only requirement is
the presence of a given set of isometries. The presence of these
isometries is clearly unaffected by adding and rescaling
these backgrounds. Similarly it is obvious from the form of the
counterterms (see
\rf{barerenorm}-\rf{rho3} below)
that if the renormalized configuration $w$ is
dualizable then so is the corresponding bare one, $W$.
For the genuine PL type dualities the situation is more subtle.
This is one of the reasons why we restrict our general considerations to
the simpler Abelian and non-Abelian dualities.

The starting point of our argument will be to assume the invariance of 
the partition function under T-duality transformation of the bare fields:
\be
Z[W;\sigma]=Z[\tilde W;\sigma]\,.
\label{baredual}
\ee
This is the relation we obtain when we go through the by now
standard gauging procedure. Actually, even
\rf{baredual} can in principle be questioned, especially in the
dimensional regularization scheme we will  employ in this paper, since
the derivation is rather formal. In particular, we know from refs. 
\cite{GRV}, \cite{AAL} 
that the gauging procedure can be afflicted by an anomaly. Though such
 questions go beyond the scope of
the present analysis, they are important and  we hope to return to them 
in future work. 

To prepare the stage, let us now consider an infinitesimal local  Weyl
transformation $\sigma(z)
\to \sigma(z) +\lambda(z)$. Equation \rf{Weyl1} can also be written
as
\be
Y[w;\sigma(z)+\lambda(z)]=Y[w+\lambda(z)\overline\beta(w);\sigma(z)]
\label{Weyl2}
\ee
Actually \rf{Weyl2} makes
sense for a local (on the worldsheet) $\lambda(z)$ only if we
consider a generalized version of the $\sigma$-model where the
background couplings are allowed to depend on the worldsheet
coordinate $z$ as in \cite{ref15}.
We can now translate \rf{Weyl2} into the language of bare quantities
using \rf{partrenorm}:
\bea
Y[w;\sigma(z)+\lambda(z)]=Z[W(w);\sigma(z)+\lambda(z)]=
Z[W(w+\lambda(z)\overline\beta(w));\sigma(z)]  
\label{renormWeyl0}\\
=Z[W(w)+\lambda(z)\frac{\partial W}{\partial w}(w)\overline
\beta(w);\sigma(z)]\,. 
\label{renormWeyl} 
\eea
{}From these last two equations we infer that
\be
Z[W;\sigma(z)+\lambda(z)]
=Z[W+\lambda(z)\overline B(W);\sigma(z)]\,,
\label{bareWeyl}
\ee
where
\be
\overline B(W)=
\frac{\partial W}{\partial w}(H(W))\overline\beta(H(W))\,.
\label{barebeta}
\ee
are the bare Weyl anomaly coefficients, and 
\be
 H(W)\equiv w
\label{H}
\ee
 defines 
the relation between bare and renormalized couplings. We are now in a position
to put  our starting equation \rf{baredual} to work. 
Applying it to \rf{bareWeyl}, we have
\bea
Z[W;\sigma(z)+\lambda(z)]=Z[W+\lambda(z)\overline
B(W);\sigma(z)]
=Z[\Gamma\big(W+\lambda(z)\overline B(W)\big);\sigma(z)] 
\label{Weyldual0}\\
=Z[\tilde W+\lambda(z)\overline\delta\Gamma(W);\sigma(z)]\,.
\label{Weyldual}
\eea
Here we have defined
\be
\lambda\overline\delta\Gamma(W)=
\Gamma\big(W+\lambda\overline B(W)\big)-\Gamma(W)+
{\cal O}\big(\lambda^2\big)\,.
\label{deltadef}
\ee
This definition makes sense since 
the space of dualizable configurations is a linear space, in the context
of Abelian or non-Abelian duality, as discussed above. 
The second of equations
 \rf{Weyldual0} deserves further comment. In the literature, 
the standard gauging
procedure which underlies \rf{baredual} was discussed only for 
couplings which do not depend explicitly on $z$, as is the case, of 
course, for the original and dual sigma models. The replacement
$W\to W+\lambda(z){\overline B}$ in \rf{Weyldual0} requires
an extension of this procedure to $z$-dependent couplings. 
This is, however, easily achieved as the only place in the standard
procedure where the $z$ dependence matters are terms where one has to
perform partial integrations. Since the only place where a partial
integration is necessary is in the Lagrange multiplier term
enforcing the flatness of the field strength, we can carry out the
functional integration over the gauge fields exactly as in the standard
case \cite{OQ,HS,JJMO}, leading to the same formal result.

Instead of performing the Weyl transformation before duality, we could 
proceed in the opposite order: 
\be
Z[W;\sigma+\lambda]=Z[\tilde W;\sigma+\lambda]=
Z[\tilde W+\lambda\overline B(\tilde W);\sigma]\,.
\label{dualWeyl}
\ee

Comparing \rf{Weyldual} and \rf{dualWeyl} thus gives
\be
\overline\delta\Gamma(W)=\overline B(\tilde W)\,.
\label{resbare}
\ee
Note that in order 
to arrive at \rf{resbare}, it was crucial to have arbitrary
functions $\lambda(z)$ at our disposal.  Had we worked with a 
simplified version
of the above line of reasoning where $\lambda$ is a constant 
parameter, we would still have \rf{resbare}, but only
up to diffeomorphisms and gauge transformations. The point is that
the partition function (in fact, already the classical action)
 is not invariant under explicitly 
$z$ - dependent diffeomorphisms and gauge transformations.
Our result \rf{resbare} is formally valid to all orders in perturbation
 theory, but
it is formulated in terms of the bare fields $W$. Its form
in the language of renormalized quantities is complicated. However,
to lowest (1 loop) order,  bare and renormalized Weyl coefficients
agree, (as can be seen from (\ref{barebeta}))
\be
\overline B(W)=\overline\beta(W)+{\cal O}(\alpha^\prime)\,.
\ee
Hence we can write to this order,
\be
\overline \delta_R \Gamma(w)=\overline \beta(w)
\ee
where $\overline\delta_R \Gamma(w)$ is defined as the renormalized
analog
of \rf{deltadef} with $W\to w, \overline B(W) \to \overline \beta(w)$.
This in turn is nothing else than  equation \rf{Haag} which we wanted to
derive.\footnote{ In fact, a slightly more precise version of it.
Since the domain of definition
of the dual mapping $\Gamma$ is not the full coupling space, but only
its
subspace under a certain set of isometries, the gradient appearing
in \rf{Haag} is not well-defined. However, the specific combination
$\overline\beta(w){\partial\over\partial w}$ is, because it represents
a directional derivative in a tangential direction - cf.
\rf{deltadef}.}

In our line of thinking, the $[\hat T,\hat R]=0$ relation has an obvious
interpretation: It expresses the fact that the form of the duality
transformation does not depend on the configuration of the worldsheet
metric $\sigma(z)$. 
In our examples, we will verify explicitly that equation \rf{Haag}
holds, providing a check of duality on the level of the Weyl anomaly
coefficients. 

\subsection{Existence of the dual model}
On the level of renormalized couplings, the naive equivalent of 
\rf{baredual} would be
\be
Y[w;\sigma]=Y[\tilde w;\sigma]\,.
\label{renormdual1}
\ee
However, it is well known that this is not correct.
For Abelian dualities \rf{renormdual1} holds at the
1-loop level \cite{HO}, 
but it is already violated at 2-loop \cite{BFHP}. Based on
these experiences it is natural to postulate that instead of
\rf{renormdual1} we have
\be
Y[w;\sigma]=Y[\tau(w);\sigma]\,,
\label{renormdual2}
\ee
where $\tau(w)$ is the quantum corrected duality transformation
\cite{Ts,BFHP,KM}
\be
\tau(w)=\tilde w+{\cal O}\Big(\alpha^\prime\Big)\,.
\label{quantumdual}
\ee
The situation is actually even more subtle since it turns out that
true quantum equivalence requires
in addition to the quantum correction \rf{quantumdual} also
a change in the renormalization scheme \cite{BFHP}.
We will not discuss
these subtleties here since we mainly concentrate on the
leading 1-loop results where they can be neglected.

Using  definition \rf{partrenorm} and assuming \rf{baredual}
we have
\be
Y[\tau(w);\sigma]=Z[W(\tau(w));\sigma]
\ee
and
\be
Y[w;\sigma]=Z[W(w);\sigma]=Z[\tilde W(w);\sigma]\,.
\ee
Quantum equivalence of the model and its dual thus follows if
\be
W(\tau(w))\approx\tilde W(w)\,,
\label{equiv}
\ee
where $\approx$ means that the equation holds up to
diffeomorphisms and gauge transformations. This can formally be
solved using \rf{H}:
\be
\tau(w)=H\big(\tilde W^{({\cal D},\Delta)}(w)\big)\,,
\ee
but it is not clear if the diffeomorphism ${\cal D}$ and the
gauge transformation $\Delta$ (both possibly infinite) can always
be chosen such that this solution is finite.
In other words, the question is whether the renormalization of the
original theory, transported by duality to its dual formulation, 
will give rise to a consistent definition of renormalized couplings
of the dual theory. We will now
 show explicitly to 1 loop order that this is the case, 
provided \rf{Haag} holds. 
Let us first discuss briefly the explicit form of the 1 loop
counterterms
(and hence the beta functions) in dimensional regularization. 
Writing 
\be
W=W(w)=w+\frac{\alpha^\prime}{\epsilon}\rho(w)+
{\cal O}\Big((\alpha^\prime)^2\Big)\,,
\label{barerenorm}
\ee
where
\bea
\rho^{(g)}_{\mu\nu}=\hat R_{(\mu\nu)}(g,b)\,,
\label{rho1}\\
\rho^{(b)}_{\mu\nu}=\hat R_{[\mu\nu]}(g,b)\,,
\label{rho2}\\
\rho^{(\phi)}=-\frac{1}{2}\, D^\mu\partial_\mu\phi-\frac{1}{24}\,b^2
\label{rho3}
\eea
with $\hat R_{\mu\nu}$ being the generalized Ricci tensor
(built from the torsion containing connection) and $b^2=b_{\mu\nu\rho}
b^{\mu\nu\rho}$ the scalar
square
of the torsion. 
The formal inverse of \rf{barerenorm} is
\be
w=H(W)=W-\frac{\alpha^\prime}{\epsilon}\rho(W)+
{\cal O}\Big((\alpha^\prime)^2\Big)\,.
\label{renormbare}
\ee
For the Weyl anomaly coefficients, one obtains
\bea
\overline\beta^{(g)}_{\mu\nu}=\beta^{(g)}_{\mu\nu}
+2D_\mu\partial_\nu\phi+{\cal O}(\alpha^\prime)\,,  
\label{betabar1}  \\
\overline\beta^{(b)}_{\mu\nu}=\beta^{(b)}_{\mu\nu}
+b_{\mu\nu} {}^\lambda\partial_\lambda\phi+
{\cal O}(\alpha^\prime)\,,  
\label{betabar2}  \\
\overline\beta^{(\phi)}=\beta^{(\phi)}
+\partial^\mu\phi\partial_\mu\phi+{\cal O}(\alpha^\prime)\,,
\label{betabar3}
\eea
where
\be
b_{\mu\nu\lambda}=\partial_\mu b_{\nu\lambda}+{\rm cyclic}\,.
\ee

Now to lowest order in $\alpha'$, we must have $\tau(w)=\tilde w \equiv
\Gamma(w)$, and therefore \rf{equiv} becomes at this level
\be
\tilde w+\frac{\alpha^\prime}{\epsilon}\rho(\tilde w)\approx
\Gamma\Big(w+\frac{\alpha^\prime}{\epsilon}\rho(w)\Big)=
\tilde w+\frac{\alpha^\prime}{\epsilon}\frac{\partial\Gamma}{\partial
w}(w)
\rho(w)\,.
\label{equiv1}
\ee
Comparing it to \rf{Haag} and using (\ref{betabar1}-\ref{betabar3})
together with the fact that
\bea
\rho^{(g)}_{\mu\nu}=\hat R_{(\mu\nu)}=\beta^{(g)}_{\mu\nu}\,,\\
\rho^{(b)}_{\mu\nu}=\hat R_{[\mu\nu]}=\beta^{(b)}_{\mu\nu}\,,   \\
\rho^{(\phi)}=-\frac{1}{2}\,D^\mu\partial_\mu\phi-\frac{1}{24}\,b^2
=\beta^{(\phi)}
\eea
we see that \rf{equiv1} is satisfied up to an infinitesimal
diffeomorphism
with parameter
\be\label{Equiv2}
\xi^\mu=\frac{\alpha^\prime}{\epsilon}\tilde g^{\mu\nu}\partial_\nu S\,,
\ee
where $S=\tilde \phi-\phi$ is the dilaton shift.

\section{The non-Abelian dual of an SU(2) model}
Let us now apply the above formalism to specific examples. 
The first model we consider is defined by the action
\be\label{nadmodel}
S=\int{\mathrm{d}}^2\sigma
\left\{
f\left(x\right)\pa_\mu x\pa^\mu x +
h\left(x\right){\mathrm{Tr}}\left[\left(g^{-1}\pa_\mu g\right)
\left(g^{-1}\pa^\mu g\right)\right]\right\}\,,
\ee
where $g\in$SU(2), $f(x)$ and $h(x)$ are arbitrary functions. 
In Eq. \rf{nadmodel} $x(\sigma )$ is a spectator field coupled to the
principal $SU(2)$ sigma model and the form of the Lagrangian follows from 
the global $SU(2)\times SU(2)$ symmetry. 
The renormalisation properties of the model (\ref{nadmodel}) are rich 
enough to test to the full our findings of the previous sections. 
\par
Using the Euler 
angles $\left(y,z,w\right)$, the parametrisation of
the SU(2) group element is 
\be
g=\exp\left(iy\tau_3\right)  
\exp\left(iz\tau_1\right)
\exp\left(iw\tau_3\right)\,\,\,,
\ee
with $\tau_a=\sigma_a/2$ and the $\sigma_a$ are 
the usual Pauli matrices.
The target spacetime metric is then given by
\be
G_{ij}=\left(
\begin{array}{cc}
f &0\\
0&hg_{ab}
\end{array}\right)\,,\quad{\mathrm{where}}\quad
{g}_{ab}=\left(
\begin{array}{ccc}
1&0&\cos(z)\\
0&1&0\\
\cos(z)&0&1
\end{array}\right)\,,
\ee
where $i,j,\dots=1,\dots,4$ label the coordinates $(x,y,z,w)$
and $a,b,\dots=2,3,4$ label the group manifold coordinates
$(y,z,w)$.
We note that for $f=1$ and $h=x^2$, the metric $G_{ij}$ 
is that of the four sphere $S^4$ and the second term
of the action coincides with the $SU(2)$ principal
chiral sigma model when $h=1$.
\par
Since the antisymmetric tensor field
and the dilaton are both absent from the model (\ref{nadmodel})
the Weyl anomaly coefficients
${\bar{\beta}}_{ij}$ are identical to the beta functions, 
$\beta_{ij}$, which are given as:
\be
{\bar{\beta}}_{ij}=\beta_{ij}= R_{ij}\,.
\ee 
\par
The non-vanishing components of the Ricci tensor are 
\bea
R_{ab}&=& \left(\delta h\right)g_{ab}\nonumber\\
\delta h&=& 
\left\{{1\over 2} 
- {1\over 2}f^{-1}\left(h'' -{1\over 2}
f^{-1}f'h'\right) -{1\over 4}\left(fh\right)^{-1}\left(h'\right)^2
\right\}\nonumber\\
R_{11}&=& \delta f\nonumber\\
\delta f&=&
{3\over 4}\left\{\left(h^{-1}h'\right)^2 - 2h^{-1}h''
+\left(fh\right)^{-1}f'h'\right\}\,,
\label{eq1}
\eea
where the prime denotes the derivative with respect to $x$.
\par
The non-Abelian dual of the original theory, (\ref{nadmodel}),
is constructed by exploiting 
the presence of the symmetry 
$g\longrightarrow LgR$, where $L$ and $R$ are constant elements of SU(2).
The classical non-Abelian dual with respect to the `left' SU(2)
symmetry 
of our model can be written (in light-cone coordinates) as:
\be
\widetilde{S}=\int{\mathrm{d}}^2\sigma 
\left\{
f\left(x\right)\pa_ +x\pa_- x +\widetilde{ M}_{ab}
\pa_+\chi^a\pa_-\chi^b\right\}\,,
\ee
where $\chi_a$ is the auxiliary field introduced in the gauging 
precedure  and 
$\pa_{\pm}=\pa_\tau\pm 
\pa_\sigma$. 
The matrix $\widetilde{M}_{ab}$ is the inverse of
\be\label{Mdef}
M^{ab}= h\delta^{ab} + f^{ab}_c\chi^c\,\,\,.
\ee
\par
The dual theory still has a `right' $SU(2)$ symmetry. To make this 
manifest we introduce spherical coordinates
\be
\chi^a=rn^a,\quad n^an^a=1,\quad n^a=(\sin\theta\cos\varphi ,
\sin\theta\sin\varphi ,\cos\theta ),
\ee 
in terms of which  
$\tilde{S}$ takes the simple form
\be\eqalign{
\tilde{S}=&
\int {\mathrm{d}}^2\sigma
\Bigl \{ f \pa_\mu x \pa^\mu x+\frac{1}{h}\pa_\mu r \pa^\mu r+
\frac{hr^2}{D}\bigl [\pa_\mu\theta\pa^\mu\theta+
\sin^2\theta\pa_\mu \varphi\pa^\mu \varphi\bigr ]\cr &-2
 \epsilon^{\mu\nu}\frac{r^3}{D}\sin
\theta\pa_\mu\theta\pa_\nu\varphi\Bigr \}}\ 
\label{eq2}
\ee
where $D=h^2+r^2$. 
To get the one loop beta functions of (\ref{eq2})
we have to compute the generalised 
Ricci tensor,  ${\widetilde{{\widehat{R}}}}_{ij}$,\footnote
{In our conventions the torsion is given by $H_{ijk}=
\pa_iB_{jk}+\pa_kB_{ij}+\pa_jB_{ki}$ and the generalised 
Riemann tensor is given by
${\widehat{R}}^i_{jkl}=\pa_j\Omega^i_{kl} + \Omega^i_{jr}\Omega^r_{kl}
-\left(k\leftrightarrow l\right)$, where $\Omega^i_{jk}=
\left\{^{\,\,i}_{jk}\right\}
+H^i_{\,\,\,jk}$ is the generalised connection. Finally, the generalised 
Ricci tensor is given by ${\widehat{R}}_{jk}={\widehat{R}}^r_{jrk}$.}
where $i,j,\dots$ label the cordinates $(x,r,\theta,\varphi)$.
The non-vanishing 
components of the generalised Ricci tensor  
are as follows 
\bea
\widetilde{\hat R}_{12}&=&\widetilde{\hat R}_{21}=
-{1\over D^2h}r{h'}(-r^2+3h^2)
\nonumber\\
\widetilde{\hat R}_{11}&=&
{1\over 4D^2h^2f}
(-r^2+3h^2)(-h^3{h'}{f'}+2h^3{h^{''}}f-3{h'}^2fh^2
+2r^2h{h^{''}}f
\nonumber\\
&-&r^2h{h'}{f'}+r^2{h'}^2f)
\nonumber\\
\widetilde{\hat R}_{22}&=&
{1\over 4D^3h^3f^2}
(-r^4{h'}^2f-r^4{h'}{f'}h+2r^4{h^{''}}hf-2r^4f^2h-2r^2{h'}{f'}h^3
+4r^2{h^{''}}h^3f
\nonumber\\
&-&12r^2h^3f^2-6r^2{h'}^2h^2f-{h'}{f'}h^5
-5{h'}^2fh^4+2{h^{''}}h^5f+6h^5f^2)
\nonumber\\
\widetilde{\hat R}_{33}&=&
{1\over 4D^3hf^2}
r^2(r^4{h'}{f'}h+r^4{h'}^2f+2r^4f^2h-2r^4{h^{''}}hf
+4r^2{h'}^2h^2f
\nonumber\\
&-&5{h'}^2fh^4+6h^5f^2+2{h^{''}}h^5f-{h'}{f'}h^5)
\nonumber\\
\widetilde{\hat R}_{34}&=&-\widetilde{\hat R}_{43}=
{r^3\sin(\theta)\over 2D^3f^2}(-r^2h{h'}{f'}-r^2{h'}^2f
+2r^2h{h^{''}}f-5{h'}^2fh^2
+2h^3{h^{''}}f
\nonumber\\
&+&4h^3f^2-h^3{h'}{f'})
\nonumber\\
\widetilde{\hat R}_{44}&=&
{r^2\sin^2\theta\over 4D^3hf^2}
(r^4{h'}{f'}h+r^4{h'}^2f+2r^4f^2h-2r^4{h^{''}}hf
\nonumber\\
&+&4r^2{h'}^2h^2f
-5{h'}^2fh^4+6h^5f^2+2{h^{''}}h^5f-{h'}{f'}h^5)
\eea
The part of the dual action involving $(r,\theta,\varphi)$,
when $h=1$, does indeed correspond to the dual of the principal 
model and if in addition $f=1$, the corresponding components
of $\widetilde{\hat R}_{ij}$ coincide also.
\par
It is worth pointing out that the one loop counterterms
do not all have the same form as those present in 
the classical dual action (\ref{eq2})
(e.g. the presence of counterterms of the form 
$\widetilde{\hat R}_{12}\pa_\mu x\pa^\mu r$).
This is to be contrasted with the original theory where all
the counterterms are of the same form as those present in
the classical Lagrangian (\ref{nadmodel}).
Consequently in the dual model one has to perform a nonlinear
renormalization of the $r$ field. 
\par
Our goal now is to show that the Weyl anomaly coefficients of the original 
and of the dual theories are related through 
the main equation of the paper.
 Equations (\ref{Haag})  
 are now explicitly written as
\bea
{\widetilde{{\bar{\beta}}}}_{(ij)}&\equiv&
{\widetilde{\beta}}_{(ij)}
+2\widetilde\nabla_i
\widetilde\nabla_j\widetilde\Phi
={\partial{\widetilde{G}}_{ij}\over\partial G_{rs}}
{\bar{\beta}}_{(rs)} + 
{\partial{\widetilde{G}}_{ij}\over\partial B_{rs}}
{\bar{\beta}}_{[rs]}
\nonumber\\
{\widetilde{{\bar{\beta}}}}_{[ij]}&\equiv&
{\widetilde{\beta}}_{[ij]}
+{\widetilde{G}}^{kl}\partial_k
{\widetilde{\Phi}}
{\widetilde{H}}_{lij}
={\partial{\widetilde{B}}_{ij}\over\partial G_{rs}}
{\bar{\beta}}_{(rs)}
+{\partial{\widetilde{B}}_{ij}\over\partial B_{rs}}
{\bar{\beta}}_{[rs]}
\eea 
Notice that $(\widetilde{G}_{ij},\widetilde{B}_{ij})$ 
depend only on the two functions
$f$ and $h$. Therefore, 
using the fact that $h={1\over 3}g^{ab}G_{ab}$ and 
$f=G_{11}$ ( that is ${\pa\over \pa G_{11}}={\pa\over\pa f}$
and ${\pa\over \pa G_{ab}}={1\over 3}g^{ab}{\pa\over \pa h}$,
where $a,b=2,3,4$),
one can write these equations as
\bea
{\widetilde{{\bar{\beta}}}}_{(ij)}&=&
{\widetilde{\beta}}_{(ij)}
+2\widetilde\nabla_i
\widetilde\nabla_j\widetilde\Phi
=\delta\Gamma_{(ij)}
\nonumber\\
{\widetilde{{\bar{\beta}}}}_{[ij]}&=&
{\widetilde{\beta}}_{[ij]}
+{\widetilde{G}}^{kl}\partial_k
{\widetilde{\Phi}}
{\widetilde{H}}_{lij}
=\delta\Gamma_{[ij]}\,\,\,,
\eea
where the non-vanishing components of the matrix $\delta\Gamma_{ij}$
are given by
\bea
\delta\Gamma_{11}&=&{\pa\widetilde{G}_{11}\over\pa f}\delta f
\nonumber\\
\delta\Gamma_{(ab)}&=&
{\partial{\widetilde{G}}_{ab}\over\partial h}
\delta h\nonumber\\
\delta\Gamma_{[ab]}&=&
{\partial{\widetilde{B}}_{ab}\over\partial h}
\delta h
\eea
In finding $\delta\Gamma_{ij}$, we have made use of the explicit
forms of $\bar{\beta}_{rs}$ in (\ref{eq1}) and of $(\widetilde{G}_{ij},
\widetilde{B}_{ij})$ as given in (\ref{eq2}).
\par 
To illustrate the self-consistency of our results 
we are now going to use the above equations {\sl to determine}
the dilaton shift $\widetilde{\Phi}$ and verify that it does coincide
with the result \cite{OQ}
$
\tilde\Phi-\Phi\sim\log\hbox{det} M\,.
$

\par
Writing $\widetilde{\Phi}$ in the general form
\be
\widetilde{\Phi} =-{1\over 2}
\ln\left[N\left(x,r,\theta,\varphi\right)\right]\,\,\,
\ee
then the equations corresponding to $\delta\Gamma_{[13]}$ and
$\delta\Gamma_{[14]}$ lead respectively to
\be
{\pa N\over \pa \theta}=0\,\,\,,\,\,\,
{\pa N\over \pa \varphi }=0\,\,\,.
\ee
Hence $N$ can only be function of $x$ and $r$, that is $N=N(x,r)$.
\par
The rest of the equations emerging from considering the 
other components of $\delta\Gamma_{ij}$, namely 
$\delta\Gamma_{(11)}$, 
$\delta\Gamma_{(12)}$, $\delta\Gamma_{(22)}$ ,
$\delta\Gamma_{(33)}$, $\delta\Gamma_{(44)}$ and 
$\delta\Gamma_{[34]}$ 
lead respectively to the following six algebro-differential equations
\bea
0&=&
(2r^4{h^{''}}hf-r^4{h'}{f'}h-2r^4{h'}^2f+8r^2{h^{''}}h^3f
-4r^2{h'}{f'}h^3-3{h'}{f'}h^5\nonumber\\
&+&6{h^{''}}h^5f-6{h'}^2fh^4)N^2
-h^2D^2(-{f'}{\pa N\over \pa x}+2f{\pa^2 N\over \pa x^2})N
+2h^2fD^2({\pa N\over \pa x})^2
\nonumber\\
0&=&-2{h'}r(-r^2+3h^2)N^2
-2hD^2N{\pa^2 N\over \pa x\pa r}
+D^2(2h{\pa N\over\pa x}-N{h'}){\pa N\over\pa r}
\nonumber\\
0&=&+2fh^3D^2({\pa N\over\pa r})^2
+(-4r^2h^3f+4fh^5-r^4{h'}^2-4{h'}^2h^2r^2-3{h'}^2h^4)N^2
\nonumber\\
&-&hD^2(2fh^2{\pa^2 N\over \pa r^2}-{h'}{\pa N\over \pa x})N
\nonumber\\
0&=&r^2(r^4{h'}^2+2{h'}^2h^2r^2+4fh^5-3{h'}^2h^4)N
+r^2{h'}h(h^2-r^2)D{\pa N\over \pa x}
\nonumber\\
&-&2fh^5Dr{\pa N\over \pa r}
\nonumber\\
0&=&
-r(r^4{h'}^2+2{h'}^2h^2r^2+4fh^5-3{h'}^2h^4)N
-r(h^2-r^2)D{\pa N\over \pa x}\nonumber\\
&+&2fh^5D{\pa N\over \pa r}
\nonumber\\
0&=&
-2r(3h^2+r^2)(hf-{h'}^2)Nud
-2r{h'}hD{\pa N\over \pa x}
+hf(3h^2+r^2)D
\eea
In these equations, we notice that ${\pa^2N\over\pa x^2}$
comes only from the $\delta\Gamma_{(11)}$ equation.
Therefore we consider this equation on its own.
On the other hand, the last five equations 
can be treated as a system of equations in
the `independent` variables
($N,{\pa N\over\pa x},{\pa N\over\pa r},
{\pa^2 N\over\pa r^2}, {\pa^2 N\over\pa x\pa r}$). 
The solution to this system is given by surprisingly
simple expressions
\be
{\pa N\over \pa r}={2Nr\over D}\,\,\,,\,\,\,
{\pa^2 N\over \pa r^2}={2N\over D}\,\,\,,\,\,\, 
{\pa N\over \pa x} = {Nh^{'}\left(3h^2+r^2\right)\over hD}\,\,\,,\,\,\,
{\pa^2 N\over \pa x\pa r} = {2rNh^{'}\over hD}\,\,.
\ee
These differential equations, in turn,  have a unique solution given by
\be
N\left(x,r\right)=ahD\,\,\,
\ee
with $D=h^2+r^2$ and $a$ a constant. This solution does
also satisfy the $\delta\Gamma_{(11)}$ equation.
\par
We note that 
\be
N\left(x,r\right)= a\det\left(M\right)\,,
\ee
where $M$ is the matrix defined in (\ref{Mdef}).
This shows that the dilaton shift $\widetilde{\Phi}$, 
determined by our conditions to ensure one loop quantum
equivalence of the dually related models 
coincides (up to the irrelevant constant $a$) with the result
of Ref.\ \cite{OQ}.

\section{Poisson Lie duality at one loop}

Let us now investigate the question of quantum T-duality for models
related by Poisson-Lie duality.
As the validity of the argument leading to \rf{Haag} is not immediately
obvious in the Poisson-Lie case, we will follow a slightly different
path as compared to the previous example. 
We will first establish that the condition of Poisson-Lie dualizability
is compatible with the 1 loop RG flow, 
at least in an infinitesimal neighborhood
of the starting point.
 We then proceed to show that \rf{equiv} can be satisfied  for a
 suitable
choice of diffeomorphism and gauge transformation. As we have seen in
section 
2.2, this equation expresses quantum T duality, and is equivalent (at
least
to 1 loop) to \rf{Haag}. 

To investigate the interplay between renormalization and 
(genuine) Poisson Lie duality we consider two mutually dual sigma models 
defined on a specific six dimensional Drinfeld double. The double we use
is the same that appeared in ref.s \cite{plex}, \cite{plex1}. 
In \cite{plex} a pair of PL 
duality related WZWN models were considered on it, while in \cite{plex1} the 
canonical transformations connecting the PL duality related family of sigma 
models were worked out. In \cite{plex1} the question of quantum equivalence of 
these sigma models was also investigated in a particular limit - when the 
target space of the sigma models reduces to two dimensions. Here we consider 
another particular example, when the target spaces of the PL dual models 
are three dimensional and investigate whether the (one loop) renormalization 
and the duality transformations commute.

The Lie algebra of the double is generated by $J_a$ and $T^a$ ($a=1,2,3$), 
satifying 
\be\label{dupla}
\eqalign{
[J_a,J_b]&=\epsilon_{abc}J_c,\qquad [T^a,T^b]=f^{ab}_{\ \ c}T^c,\cr
[J_a,T^b]&=f^{bc}_{\ \ a}J_c-\epsilon_{acb}T^c}
\ee
where the only non vanishing commutator among the $T^a$-s are 
$[T^i,T^3]=cT^i$ for $i=1,2$ with $c$ being a constant. (If $c\ne 0$ then it
could be absorbed into $T^3$, but we keep it, as then seting $c=0$ should
lead back to the case of ordinary non-Abelian duality).  The $T^a$
generators may be thought of as describing two translations and a dilatation. 
It is straightforward to determine the various matrices of 
the Klim\v c\'\i k -- Severa construction  
using e.g. the $6\times 6$ representation of the generators
following from Eq.(\ref{dupla}). 
Parametrizing the elements of the $SU(2)$ subgroup as: 
\be
g=\exp(J_3f)\exp(J_2t)\exp(J_3p),
\ee
the $b_{ab}(g)$, $d_{ab}(g)$ matrices are defined as:
\be
g^{-1}T^ag=b_{ab}(g)J_b+d_{ab}(g)T^b,
\ee
while the third matrix of this construction, 
$g^{-1}J_ag=a_{ab}(g)J_b$, 
is related to $d_{ab}$ 
by
 $d^t_{ab}(g)=a^{-1}_{ab}(g)$. 
This can be derived from the consistency of the 
invariant scalar product on the double:
\be
<J_a,J_b>=0,\quad <T^a,T^b>=0,\quad <J_a,T^b>=\delta_a^b,
\ee

{}From the explicit form of these matrices we find the $\Pi$ matrix as:
\be
\Pi=b(g)a^{-1}(g)=
\left [\begin {array}{ccc} 0&-c\cos(t)+c&\sin(f)c\sin(t)
\\\noalign{\medskip}c\cos(t)-c&0&-\cos(f)c\sin(t)\\\noalign{\medskip}-
\sin(f)c\sin(t)&\cos(f)c\sin(t)&0\end {array}\right ].
\ee
The Lagrangian of the original model is given by
\be\label{lag}
{\cal L}=(E+\Pi )^{-1}_{ij}(\pa_+g\, g^{-1})_i(\pa_-g\, g^{-1})_j
\ee
where $E$ is an arbitrary constant matrix and $(\pa_+g\, g^{-1})_i$ are the
right invariant one forms:  
$(\pa_+g\, g^{-1})_i={\rm Tr}(\pa_+g\, g^{-1}J_i)$. Various choices of $E$ 
give different (dualizable) $\sigma $ models living on the same double, and 
the expression in Eq.(\ref{lag}) is the most general solution of the 
dualizability conditions. 
Since 
both  $(\pa_\pm g\,g^{-1})_i$ and $\Pi $ depend only on the derivatives of  
the $p$ field but not on $p$ itself, ${\cal L}$
is invariant under the $p$ translations for 
any choice  of $E$. The simplest choice of the original model is to 
set $E\equiv {\rm Id}$, and in this case we find ($t_\pm\equiv\pa_\pm t$ etc.)
\be\label{explag}\eqalign{
&{\cal L}= 
\frac {16}{-1+2\,c^{2}\cos(t)-2\,c^{2}}
\Bigl [
-{\it t_+}\,{\it t_-}+{\it p_+}\,c^{2}{\it f_-}-{\it f_+}\,c^{2}{\it f_-}-{
\it p_+}\,{\it p_-}-{\it p_+}\,c^{2}{\it p_-}\cr &
-{\it f_+}\,{\it f_-}+{\it f_+}
\,c^{2}{\it p_-}
-c\left (-{\it t_-}\,{\it f_+}+{\it t_+}\,{\it p_-}-{\it t_-
}\,{\it p_+}+{\it t_+}\,{\it f_-}\right )\sin(t)\cr &
+\left (2\,{\it 
p_+}\,c^{2}{\it p_-}+2\,{\it f_+}\,c^{2}{\it f_-}-2\,{\it p_+}\,c^{2}{\it 
f_-}-2\,{\it f_+}\,c^{2}{\it p_-}-\,{\it p_+}\,{\it f_-}-\,{\it f_+}\,{
\it p_-}\right )\cos(t)\cr &
-c^{2}\left (-{\it f_-}+{\it p_-}\right )
\left (-{\it f_+}+{\it p_+}\right )\cos(t)^{2}\Bigr ]\ .}
\ee
Note that ${\cal L}$
is also invariant under the $f$ translations. Writing 
\be
{\cal L}=g_{\mu\nu}\pa_i \xi^\mu\pa^i \xi^\nu+
\epsilon_{ij}B_{\mu\nu}\pa^i \xi^\mu\pa^j \xi^\nu
\ee
($\xi^\mu=t,\ f,\ p$) 
effectively means that we identify the metric as the coefficients of 
$(\xi^\mu_+\xi^\nu_-+\xi^\nu_+\xi^\mu_-)/2$ and the torsion potential, 
$B_{\mu\nu}$, as 
the coefficients of $(\xi^\mu_+\xi^\nu_--\xi^\nu_+\xi^\mu_-)/2$ in 
Eq.(\ref{explag}).
It turns out that this $B_{\mu\nu}$ is a pure gauge, as it has vanishing 
torsion.
Therefore the one loop counterterm, $\Sigma_1$, is determined by 
the (ordinary) Ricci tensor, $R_{\mu\nu}$, which is {\sl symmetric} in 
$\mu ,\nu$.

 Since Eq.(\ref{lag}) (with an arbitrary $E$) is the most general 
solution of the dualizability condition we can go on with this construction
if the change in the background caused by renormalization can be absorbed by
changing the $E$ matrix. Thus 
the -- one loop --  compatibility between Poisson-Lie dualizabilability and
renormalization requires
\be\label{compat}
\alpha^\prime R_{\mu\nu}\pa_+\xi^\mu\pa_-\xi^\nu=\alpha^\prime 
\frac{\pa{\cal L}}{\pa n_k} Y_k(n),
\ee
where the $n_k$ ($k=1,..9$) parametrize the $E=E(n_k)$ matrix (e.g. $n_k$ are 
its elements) and  change as $n_k\rightarrow n_k+\alpha^\prime Y_k(n)$ under 
renormalization. 
In our case $E(n_k)\vert_{n=0}={\rm Id}$ thus the right hand
side of Eq.(\ref{compat}) becomes
\be\label{compat1}
-\Bigl[ ({\rm Id}+\Pi )^{-1}\frac{\pa E(n)}{\pa n_k}Y_k(n)\vert_{n=0}
({\rm Id}+\Pi )^{-1}\Bigr]_{ij}(\pa_+g\, g^{-1})_i(\pa_-g\, g^{-1})_j
\ee
where $S\equiv\frac{\pa E(n)}{\pa n_k}Y_k(n)\vert_{n=0}$ is {\sl a completely 
arbitrary} $3\times 3$ matrix. Thus in our case the compatibility between 
the renormalization flow and the dualizability condition depends on whether we can 
choose the (a priori completely arbitrary) elements of $S$ in such a way that
\be\label{compat2}\eqalign{
R_{\mu\nu}\pa_+\xi^\mu\pa_-\xi^\nu &=  
-\Bigl[ ({\rm Id}+\Pi )^{-1}S
({\rm Id}+\Pi )^{-1}\Bigr]_{ij}(\pa_+g\, g^{-1})_i(\pa_-g\, g^{-1})_j
\cr &\equiv M_{\mu\nu}\pa_+\xi^\mu\pa_-\xi^\nu}\ ,
\ee
holds, where the left hand side is given by the specific $R_{\mu\nu}$ following 
from Eq.(\ref{explag}). More precisely the {\sl symmetric} part of $M_{\mu\nu}$ 
must coincide with $R_{\mu\nu}$, while for its {\sl antisymmetric} part it is
enough to require a vanishing torsion:\footnote{Since $t$, $f$ and $p$ have
a geometrical meaning (i.e. they are angular variables on $S^3$) we do not 
expect any field renormalizations (reparametrizations) for them, that's why 
these terms are missing in Eq.(\ref{compat}).}
\be
\partial_p(M_{[12]})+\partial_t(M_{[23]})+\partial_f(M_{[31]})=0.
\ee 
These requirements lead only to $6+1=7$ 
equations for the $9$ unknown elements of $S$. 
Writing 
\be
S=\pmatrix{m_1&m_2&m_3\cr m_4&m_5&m_6\cr m_7&m_8&m_9}
\ee
Maple found from these seven equations that all the other $m_k$ can be  
determineded in terms of $m_2$ and $m_4$ if they are related as $m_4=-m_2$. 
However,
looking at the explicit solution 
 we deduce that the 
$m_k$ obtained this way are constant (i.e. are independent of $t$, $f$ and
$p$) only for $m_2=c$, and for these values $S$ takes 
the form:
\be
S=\pmatrix{\frac{1}{2}&c&0\cr -c&\frac{1}{2}&0\cr 
                      0&0&2c^2+\frac{1}{2}}.
\ee   
Thus -- for the simplest original model at least --  the
dualizability condition seems to be satisfied along the renormalization flow, 
and this allows us to construct the righthand side of \rf{equiv}.
The validity of \rf{equiv} then means that the dualization of the model
with bare couplings of the form \rf{barerenorm} will lead to a dual theory
which is finite in the limit $\epsilon \to 0$, for a suitable
choice of the diffeomorphism and gauge transformation implicit in \rf{equiv}. 
In order to see this, we will study the renormalization of the dual model
and check that the required renormalizations of the dual couplings are exactly
those induced from the original model by the duality transformation.

According to the Klim\v c\'\i k-Severa construction the dual Lagrangian is again of
 the same form as in \rf{lag}:
\be
{\tilde {\cal L}}=(\tilde E+\tilde \Pi )^{-1}_{ij}
(\pa_+\tilde g\, {\tilde g}^{-1})_i(\pa_-\tilde g\,{\tilde g}^{-1})_j
\label{duallag}
\ee
The  constant matrix $\tilde{E}$
 is related to $E$ by $E\tilde{E}={\rm Id}$, and this should be true even
when $E$ undergoes a renormalization as above: 
\be\label{etilde}
\tilde{E}\equiv {\rm Id}+\alpha^\prime \tilde{S}=E^{-1}=
\Bigl({\rm Id}+\alpha^\prime S\Bigr)^{-1}={\rm Id}-\alpha^\prime S+...\ .
\ee
On the other hand, $\tilde S$ will be determined independently 
by the analog of the above
renormalization analysis for the dual model, and thus \rf{etilde}
is a nontrivial condition. 

Parametrizing the elements of the dual group as
\be\label{dualisp}
\tilde{g}=\exp(XT^1)\exp(YT^2)\exp(ZT^3),
\ee
we find the dual equivalent of the $\Pi $ matrix in the form:
\be
 \tilde{\Pi }=   
\left [\begin {array}{ccc} 0&{\frac {X^{2}c^{2}+Y^{2}c^{2}+{e^{-2\,Zc}
}-1}{2\,c}}&Y\\\noalign{\medskip}-{\frac {X^{2}c^{2}+Y^{2}c^{2}+{e^{-2
\,Zc}}-1}{2\,c}}&0&-X\\\noalign{\medskip}-Y&X&0\end {array}\right ].
\ee
(Note that for $c\rightarrow 0$ $\tilde{\Pi }$ reduces to what we already
know from the non-Abelian dual of the principal model). Using 
Eq.(\ref{dualisp}) the coefficients of the right invariant one form 
$\pa_\pm \tilde{g} \tilde{g}^{-1}=(\pa_\pm \tilde{g} \tilde{g}^{-1})_iT^i$ 
are given by very simple expressions:
\be
(\pa_\pm \tilde{g} \tilde{g}^{-1})_1=X_\pm +cXZ_\pm ,\quad
(\pa_\pm \tilde{g} \tilde{g}^{-1})_2=Y_\pm +cYZ_\pm ,\quad
(\pa_\pm \tilde{g} \tilde{g}^{-1})_3=Z_\pm .
\ee

The dual Lagrangian, following from the dual version of Eq.(\ref{lag}) 
(with $\tilde{E}={\rm Id}$), becomes easily tractable if we use 
cylindrical coordinates:
\be
X=r\cos (\alpha ),\qquad Y=r\sin (\alpha ),\qquad Z=z.
\ee
Indeed  introducing the notation 
$\chi^1=r$, $\chi^2=\alpha $, $\chi^3=z$, and defining
\be
D=
c^{4}r^{4}+2\,c^{2}r^{2}\left (1+{e^{-2\,cz}}\right )+4\,c^{2}+\left (
1-{e^{-2\,cz}}\right )^{2},
\ee
together with 
\be
M=
r^{2}c^{2}+2\,c^{2}+1-{e^{-2\,cz}},
\ee
the Lagrangian, $\tilde{\cal L}=\tilde{g}_{\mu\nu}\pa_i\chi^\mu\pa^i\chi^\nu +
\tilde{b}_{\mu\nu}\epsilon_{ij}\pa^i\chi^\mu\pa^j\chi^\nu$, assumes the form:
\be
\eqalign{
\tilde{\cal L}&=
\frac{1}{D} \Bigl[ 4\,c^2(1+r^2)\pa_i r^2
+4\,c^2r^2\pa_i  
\alpha ^2+\bigl (4\,c^2(e^{-2\,cz}-c^2)+M^2
\bigr )\pa_i z^2\cr &
+4\,cr
M\pa_i r\,\pa^i z
+
2\,\epsilon_{ij}\,\bigl (-2\,cr\left (-1+e^{-2\,cz}+r^2c^2\right )
\pa^i r\,\pa^j\alpha
\cr &+2\,c^2r^2\left (1+e^{-2\,cz}+r^2c^2\right )
 \pa^i\alpha\,\pa^j z \bigr )\Bigr ]\ .}
\ee
Note that this Lagrangian is independent of $\alpha $.  The metric appearing
here has only four non vanishing components while the antisymmetric 
tensor field, 
$b_{\mu\nu}$, has two non vanishing components. 
The one loop counterterm is determined by the generalized 
Ricci tensor. The explicit form of $\widetilde{\hat{ R}}_{\mu\nu}$
 reveals that 
$\widetilde{\hat{R}}_{12}=-\widetilde{\hat{R}}_{21}$, 
$\widetilde{\hat{R}}_{23}=-\widetilde{\hat{R}}_{32}$, 
$\widetilde{\hat{R}}_{13}=\widetilde{\hat{R}}_{31}$, i.e.  
the one loop counterterm containes no new 
derivative couplings that are not present in $\tilde{\cal L}$. 

In the dual model we have to check whether the dual version of the 
renormalizability 
condition, Eq.(\ref{compat}),
\be\label{dcompat}
\alpha^\prime \widetilde{\hat{R}}_{\mu\nu}\pa_+\chi^\mu\pa_-\chi^\nu=
\alpha^\prime 
\frac{\pa\tilde{{\cal L}}}{\pa \tilde{n}_k} \tilde{Y}_k(\tilde{n})+
\alpha^\prime\frac{\delta\tilde{{\cal L}}}
{\delta\chi^\mu}\chi^\mu_1(\chi^\nu),
\ee
is satisfied by the $\tilde{E}$ in Eq.(\ref{etilde}). In this equation 
$\chi^\mu_1(\chi^\nu)$ denote the potential 
re\-pa\-ra\-met\-ri\-za\-tions (which we expect,
since in the dual of the principal model they were also present), and as
usual, equality is required modulo the gauge transformations on the 
antisymmetric parts.  

{}From the (anti)symmetry properties of $\widetilde{\hat{R}}_{\mu\nu}$ -- by 
the same arguments as in the paper on NAD, Bal\'azs et al. \cite{ex},  
 -- we conclude that the
 reparametrizations must have the form:
\be
r_0\rightarrow r+\frac{\alpha^\prime}{\epsilon}F(r,z),\qquad
z_0\rightarrow z+\frac{\alpha^\prime}{\epsilon}G(r,z),\qquad
\alpha_0 \rightarrow \alpha +\frac{\alpha^\prime}{\epsilon}y_2\alpha ,
\ee
with $y_2$ being an $r$ and $z$ independent constant. Therefore the last 
term in Eq.(\ref{dcompat}) is independent of the $\alpha $ field. The first
term in  Eq.(\ref{dcompat}) is computed in the same way as in
 the case of the 
original model, Eq.(\ref{compat1}): denoting 
$\frac{\pa \tilde{E}(\tilde{n})}{\pa \tilde{n}_k}
\tilde{Y}_k(\tilde{n})\vert_{\tilde{n}=0}\equiv\tilde{S}$, it has the form: 
\be\label{dcompat2}
-\Bigl[ ({\rm Id}+\tilde{\Pi} )^{-1}\tilde{S}
({\rm Id}+\tilde{\Pi} )^{-1}\Bigr]_{ij}
(\pa_+\tilde{g}\, \tilde{g}^{-1})_i(\pa_-\tilde{g}\, \tilde{g}^{-1})_j
\equiv \tilde{M}_{\mu\nu}\pa_+\chi^\mu\pa_-\chi^\nu.
\ee
A hopeful sign of consistency comes from the observation that 
$\tilde{M}_{\mu\nu}$ becomes $\alpha $ independent (which we need as the other
two pieces in Eq.(\ref{dcompat}) are also $\alpha $ independent), if 
$\tilde{S}$ has the form:
\be
\tilde{S}=\pmatrix{\tilde{m_1}&\tilde{m_2}&0\cr 
                   -\tilde{m_2}&\tilde{m_1}&0\cr
                   0&0&\tilde{m_9}},
\ee
with arbitrary $\tilde{m_1}$, $\tilde{m_2}$ and $\tilde{m_9}$. Furthermore 
this choice of $\tilde{S}$ yields an $\tilde{M}$ that also satisfies 
$\tilde{M}_{12}=-\tilde{M}_{21}$, $\tilde{M}_{32}=-\tilde{M}_{23}$, 
$\tilde{M}_{13}=\tilde{M}_{31}$. Therefore we use this $\tilde{M}_{\mu\nu}$ 
in Eq.(\ref{dcompat}). It is straightforward to obtain the 
{\sl symmetric} part
of Eq.(\ref{dcompat}), i.e. the equalities of the coefficients of 
$(\pa_i r )^2$, $(\pa_i\alpha )^2$, $(\pa_i z)^2$ and  
$\pa_i r\pa^i z$ on the two sides. Denoting by $(LS)_{\mu\nu}$,
 $(LA)_{\mu\nu}$ the symmetric and antisymmetric 
coefficients of the field derivatives in 
$\frac{\delta\tilde{{\cal L}}}
{\delta\chi^\mu}\chi^\mu_1(\chi^\nu)$:
\be
\frac{\delta\tilde{{\cal L}}}
{\delta\chi^\mu}\chi^\mu_1(\chi^\nu)=
(LS)_{\mu\nu}\pa_i\chi^\mu\pa^i\chi^\nu +
(LA)_{\mu\nu}\epsilon_{ij}\pa^i\chi^\mu\pa^j\chi^\nu,
\ee
and introducing the notation\footnote{Note that $FR,FZ,GR,GZ$ are single
variables and should not be read as products.}:
\be\label{ddef}
\pa_r F=FR,\quad \pa_z F=FZ,\quad \pa_r G=GR,\quad \pa_z G=GZ, 
\ee
the symmetric part of Eq.(\ref{dcompat}) can be written as:
\be\label{ddeq}
(LS)_{\mu\nu}(F,G,FR,FZ,GR,GZ)
=\widetilde{\hat{R}}_{\mu\nu}-\tilde{M}_{\mu\nu},\qquad 
(\mu\, \nu =11,\ 22,\ 33,\ 13).
\ee 
(Note that the left hand side depends linearly on $F$, $G$,
$FR$, $FZ$, $GR$ and $GZ$). 
We handle the {\sl antisymmetric} part of Eq.(\ref{dcompat}) in the following
way: we define $\Delta^{12}$ and $\Delta^{23}$ as the coefficients of 
$\epsilon_{ij}\pa^i r\pa^j\alpha $ (respectively 
$\epsilon_{ij}\pa^i \alpha \pa^j z$) in the {\sl difference} 
between the left hand sides and the right hand sides of Eq.(\ref{dcompat}). 
Then the requirement that the two sides of Eq.(\ref{dcompat}) be equal up
to a gauge transformation can be written as:
\be\label{tors}
\pa_z(\Delta^{12})+\pa_r(\Delta^{23})=0.
\ee
Naively this equation containes second derivatives of $F(r,z)$ and $G(r,z)$, 
however it is easy to show, that the antisymmetry of $\tilde{b}_{\mu\nu}$ 
guarantees that all second derivatives of $F$ and $G$ cancel from 
Eq.(\ref{tors}). 
Therefore Eq.(\ref{ddeq}) and 
Eq.(\ref{tors}) give five (linear) equations for the six unkowns $F$, $G$,
$FR$, $FZ$, $GR$ and $GZ$. The great surprise is that $F$ and $G$ can be 
expressed {\sl completely algebraically} from these equations, while $FR$, 
$GR$ and $FZ$ are given by three (inhomogeneous) linear expressions in $GZ$. 
Renormalizability of the dual model amounts to the problem of 
choosing the $y_2$, $\tilde{m}_1$, $\tilde{m}_2$ and $\tilde{m}_9$ 
parameters in such a way (if there is any), that the $r$ and $z$ derivatives 
of the algebraically determined $F$ and $G$ functions satisfy the 
aforementioned three linear equations between $FR$, $GR$, $FZ$ and $GZ$. The 
claim is that
\be
y_2=0,\qquad 
\tilde{S}=-\pmatrix{\frac{1}{2}&c&0\cr -c&\frac{1}{2}&0\cr 
                      0&0&2c^2+\frac{1}{2}},
\ee
is the only such choice. In this case, 
 $F$ and $G$ simplify to                      
\be\label{ffu}
F=
\frac {c^2\,r}{D}\left (P^{2}+4\,P-1-4\,c^{2}+2\,r^{2}c^{2}P+c^
{4}r^{4}\right ) ,
\ee
\be\label{gfu}
G=
-\frac {c}{D}\,\left (P^{2}-1-4\,c^{2}-4\,r^{2}c^{2}+2\,r^{2}c^{2
}P+c^{4}r^{4}\right ) ,
\ee
where $P={\rm e}^{-2cz}$.

Thus $\tilde S=-S$ as required by \rf{etilde}, and we conclude that
 quantum T duality
 holds for our PL example. 
So we see that 
-- at least for $E={\rm Id}$ -- the infinitesimal renormalization flow and 
the PL duality transformations commute in the same way as for the Abelian 
and non-Abelian cases.  
The $F$ and $G$ reparametrizations 
appearing in Eq.(\ref{ffu}-\ref{gfu}) have some interesting properties. For 
$c\rightarrow 0$ they give
\be
F\rightarrow \frac{\, r}{1+r^2+z^2},\qquad 
G\rightarrow \frac{\, z}{1+r^2+z^2},
\ee
which is: (I) identical to what was found for the NAD of the principal model 
(of course {\sl without} the rescaling by the coupling constant), (II) 
identical to what is obtained from the \lq dilaton shift' 
for the $c\equiv 0$ 
(NAD) case (see Eq.\ (\ref{Equiv2})). Tyurin and Unge, \cite{TU}, 
 gave an expression for the (dual) dilaton shift 
in the general  
case; for our model ($c\ne 0$) this dilaton shift is proportional to 
$\ln D$. Note, however, that the appropriate gradient of this, 
$\frac{1}{2}\tilde{g}^{ij}\pa_j\ln D$, gives expressions, which are 
{\sl different} from those in Eq.(\ref{ffu}-\ref{gfu}). Nevertheless these
$F$ and $G$ can also be written as the components of a gradient, namely as 
$\frac{1}{2}\tilde{g}^{ij}\pa_j\ln (D{\rm e}^{2cz})$. 
This may be interpreted in two
ways: either that for PL duality the relation between the reparametrization 
and the dilaton shift is different from the 
one found for the Abelian and non-Abelian dualities, 
or that we are working in a coordinate system, which is 
not the \lq natural' one (i.e. where the connection between the dilaton 
shift and the reparametrizations is the simple gradient form).     
Also an interesting 
problem is that our $\tilde{f}^{ab}_{\ \ c}\equiv f^{ab}_{\ \ c}$ have non 
vanishing traces $f^{31}_{\ \ 1}=f^{32}_{\ \ 2}\ne 0$ -- when Tyurin and Unge 
 found that quantum equivalence is broken already in the conformal case -- 
yet we found this equivalence in the form of the commutative nature of the 
(one loop) renormalization flow and the PL duality transformations 
-- at least at a certain point of the modulus space described by $E={\rm Id}$.

\section{Conclusions and discussion}

In this paper we studied the question of quantum equivalence among dually
related sigma models in perturbation theory. Using the anomalous Ward-identity 
for Weyl symmetry, we derived the Haagensen-Olsen \cite{HO}
relation between the  Weyl-anomaly 
coefficients (and beta functions) of these models and their duals in a
 general context of Abelian or non-Abelian dualities.
We obtained this relation (to 1-loop order in terms of renormalized
quantities) from the assumption that the standard gauging 
procedure is valid for the bare fields. We pointed out that it also
implies the quantum equivalence of the dual models in the sense that
duality induces the correct renormalization of the dual model from
the renormalization of the original one. 

This simple criterion - which we believe to hold even in the 
PL context -  certainly provides a convenient way to check
quantum T duality at an elementary level, 
as we illustrated through the study of a non-Abelian
example (based on $SU(2)$) with a single spectator field.

On the other hand, it is clear that a
comprehensive analysis must go further than this, for several reasons:
As T duality should be checked for physical quantities and not just for the
partition function, sources should be introduced. Second, the question of the
existence of
a consistent regularization compatible with duality should be addressed;
dimensional regularization, while formally preserving duality, has well-known
problems with the antisymmetric tensor coupling (though this should be 
irrelevant at the one-loop level we have been considering). 
Finally, an extension of our general argument to all orders in loops 
seems desirable.
The reason why we restricted our attention to the 1-loop order 
 is that the 2 (and higher) loop problems are not only
 technically more involved, but  also conceptionally different
from the 1-loop case because of the quantum modifications of both
the transformation formulae and the renormalization \lq scheme' 
\cite{BFHP}-\cite{ex}. 

At the two-loop level,  up to now the situation is clear  only
in the Abelian case \cite{KM} where the corrections
to the classical duality transformation have been obtained in general. 
The derivation of these corrections for the non-Abelian case would be 
interesting, since no special non-Abelian example with two loop 
equivalence is known at present. 

Moreover much work remains to be done in the larger context of PL duality. 
 Our general considerations do not immediately apply for the case of genuine
Poisson-Lie dualities, due to the problem that PL dualizability - not 
being related to any isometry in general - 
cannot as easily be  seen to survive renormalization as in the traditional
Abelian and non-Abelian cases. For the  simple PL example we studied, however, 
it turned out that dualizability is indeed compatible with renormalization
at 1 loop,  
 and quantum equivalence could be established on the same level as for the
non-Abelian example.

This finding, when taken together with the one loop equivalence found for 
sigma models with two dimensional target spaces \cite{plex1}, and the
quantum equivalence of the special WZNW models described in 
Ref.\ \cite{plex}, 
gives some support to the expectation that quantum PL T-duality 
exists in a form similar to the Abelian and non-Abelian cases. For 
PL dualities, however, even the question of one-loop equivalence
has not been settled.
The conformal example of Ref.\ \cite{plex} and the non conformally invariant  
ones of Ref.\ \cite{plex1} and of section 4
are defined on a Drinfeld double which does not
satisfy the conditions which, according to  Ref.\ \cite{TU}, guarantee that the
dual theory is conformally invariant if the original one is. 
Obviously more work is needed to clarify this point.

\vskip1truecm
\noindent
{\bf Acknowledgements}

This investigation was supported in part by the Hungarian National
Science Fund (OTKA) under T016251 and T019917.

\end{document}